\documentclass[10pt,letterpaper]{article}
\usepackage[top=0.85in,left=2.75in,footskip=0.75in]{geometry}

\usepackage{amsmath,amssymb}

\usepackage{changepage}

\usepackage{textcomp,marvosym}

\usepackage{cite}

\usepackage{nameref,hyperref}

\usepackage[right]{lineno}

\usepackage[nopatch=eqnum]{microtype}
\DisableLigatures[f]{encoding = *, family = * }

\usepackage[table]{xcolor}

\usepackage{array}

\newcolumntype{+}{!{\vrule width 2pt}}

\newlength\savedwidth

\raggedright
\usepackage[aboveskip=1pt,labelfont=bf,labelsep=period,justification=raggedright,singlelinecheck=off]{caption}

\usepackage{float}

\makeatletter
\renewcommand{\@biblabel}[1]{\quad#1.}
\makeatother

\usepackage{lastpage,fancyhdr,graphicx}
\usepackage{epstopdf}
\fancyheadoffset[L]{2.25in}
\fancyfootoffset[L]{2.25in}
\begin{document}
\vspace*{0.2in}

\begin{flushleft}
{\Large
\textbf\newline{Glass-based physical models for tissue mechanics} 
}
\newline
\\
Gopika Madhu\textsuperscript{1\Yinyang},
Carolyn Delli-Santi\textsuperscript{1\Yinyang},
Jenna Efrein\textsuperscript{2},
Prannoy Suraneni\textsuperscript{3},
Landolf Rhode-Barbarigos\textsuperscript{3},
Vivek N. Prakash\textsuperscript{1,4,5*}
\\
\bigskip
\textbf{1}{Department of Physics, College of Arts and Sciences, University of Miami, Coral Gables, FL 33146, U.S.A.}
\\
\textbf{2}{Department of Art and Art History, College of Arts and Sciences, University of Miami, Coral Gables, FL 33146, U.S.A.}
\\
\textbf{3}{Department of Civil and Architectural Engineering, College of Engineering, University of Miami, Coral Gables, FL 33146, U.S.A.}
\\
\textbf{4}{Department of Biology, College of Arts and Sciences, University of Miami, Coral Gables, FL 33146, U.S.A.}
\\
\textbf{5}{Department of Marine Biology and Ecology, Rosenstiel School of Marine, Atmospheric and Earth Science, University of Miami, Miami, FL 33149, U.S.A.}
\\
\bigskip

%
%
\Yinyang These authors contributed equally to this work.





*Corresponding author, E-mail: vprakash@miami.edu

\end{flushleft}
\section*{Abstract}
Techniques from glass art and fabrication provide a controllable physical platform for studying tissue mechanics in simple organisms. Here, we use glass-based physical models to investigate tissue deformation in the marine organism \textit{Trichoplax adhaerens}. Previous studies have shown that the epithelial tissues in \textit{T. adhaerens} undergo large deformations and form fracture holes under mechanical loading, exhibiting a ductile-to-brittle transition at fast loading rates. To model these behaviors in a tunable and experimentally accessible system, glass is shaped into tissue-like monolayers in a glass studio, heated to its specific process temperature, and subjected to controlled stretching. Rapid cooling arrests the deformed configurations, providing snapshots of tissue-like strain states under load. Under lateral and radial stretching, we quantify changes in the area and eccentricity of individual “cells” in the glass models, and found that eccentricity increases after stretching. We further use tensegrity-based models to quantify deformations in the cellular geometry of the glass tissues, enabling direct comparison between experiments and simulations. The model captures the principal experimental deformation patterns, but underestimates the magnitude of the observed eccentricity changes. Our results demonstrate that glass-based physical models provide an experimentally accessible platform for studying tissue-scale deformation and mechanical behavior, while supporting interdisciplinary approaches that connect methods in the arts and sciences.



\section*{Introduction}

\subsection*{Tissue Mechanics and Epithelial Deformation}
Tissue mechanics plays a central role in regulating cell growth, proliferation, migration, function, and differentiation~\cite{Fung1990,Ayad_2019,guillot2013}. In multicellular organisms, epithelial tissues act as protective and functional barriers and are continuously subjected to mechanical forces arising from both their environment and internal cellular activity. Understanding the mechanical properties of epithelial tissues is therefore essential for linking cellular interactions to tissue-scale behavior~\cite{Fung1990}. The simplest epithelial systems are one-cell-thick monolayers, which provide experimentally tractable models for biophysical investigation~\cite{harris12,park2015unjamming}. Notably, the stress measured at fracture sites within epithelial monolayers has been estimated to be approximately nine times larger than the stress required to separate isolated cells~\cite{harris12}. Despite the prevalence of epithelial monolayers and their importance in development and physiology, experimentally accessible and intuitive methods for characterizing their mechanical properties remain limited, motivating the development of alternative experimental approaches~\cite{harris12}.

\subsection*{\textit{Trichoplax adhaerens} as a Model System}
The model system used here to study epithelial tissue mechanics is the marine organism \textit{Trichoplax adhaerens}. This placozoan is found in benthic ecosystems across all major oceans~\cite{pearse07}. It is a flat organism, approximately 25~\textmu m thick and a few millimeters in lateral extent~\cite{prakash2021motility}. It is composed of only six cell types organized into a simple three-layer body plan, consisting of an upper and a lower epithelium separated by a layer of fiber cells~\cite{SMITH2014,Armon18}. The organism feeds by external digestion through the ventral epithelium and lacks organs or muscles~\cite{smith2015}. The organism glides on substrates by generating traction forces via adhesive cilia present on the ventral epithelium~\cite{madhu2024}. 

\begin{figure}[H]
    \begin{center}
        \includegraphics[width=0.98\textwidth]{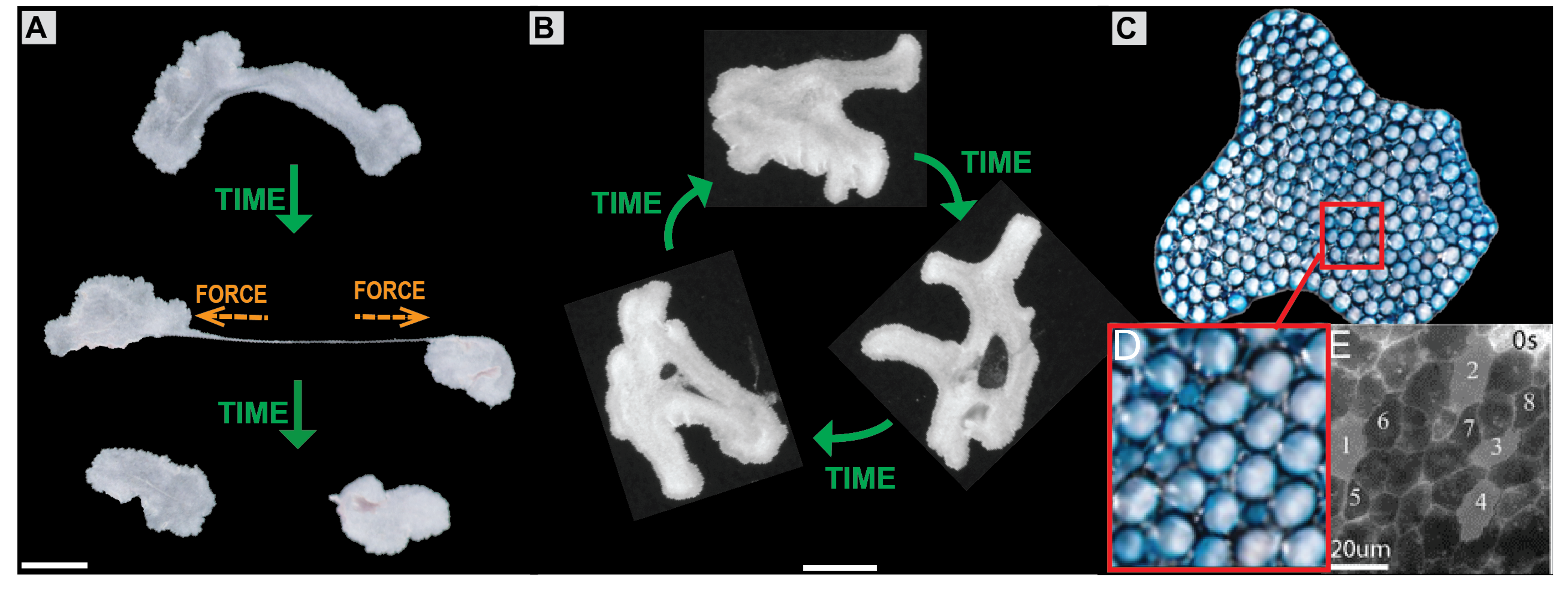}
    \caption{\textbf{Tissue deformation in \textit{Trichoplax adhaerens} and glass “tissue” models.} (A) Asexual reproduction in \textit{T. adhaerens} occurs through ductile deformation (tissue thinning) followed by binary fission. (B) Tissue fracture and subsequent healing in \textit{T. adhaerens}. Images in A and B are adapted and reproduced from reference~\cite{prakash2021motility}. (C) Glass ``tissue'' monolayer. Inset (D) shows zoom-in views of individual ``cells'' in the glass epithelium, and Inset (E) shows cells in the epithelium of \textit{T. adhaerens} (Image adapted and reproduced from reference~\cite{Armon18}). All scale bars (except C) correspond to 1 mm.}
    \label{Fig1}
    \end{center}
\end{figure}

Reproduction in \textit{T. adhaerens} occurs through ductility-induced ruptures, effectively a form of binary fission~\cite{prakash2021motility}. Because growth and division are governed by mechanical failure rather than by a fixed developmental program, these organisms do not possess a defined body shape~\cite{prakash2021motility}. Instead, they continuously deform, migrate, and split apart, providing a direct experimental window into the limits of epithelial tensile strength (Fig.~\ref{Fig1}). When viewed through the lens of material behavior, the tissue dynamics of \textit{T. adhaerens} exhibit elastic, ductile, and brittle regimes, making it a useful system for studying tissue mechanics~\cite{gooshvar2023}. For example, the formation of thin cellular threads resembles ductile deformation, whereas tissue rupture corresponds to brittle fracture~\cite{prakash2021motility}. Computer simulations of this system using a simple adhesive ball-and-spring model revealed an elastic–ductile–brittle transition driven by local cellular rearrangements~\cite{prakash2021motility}. 

\subsection*{Parallels Between Tissues and Glassy Materials}
Mechanical deformation of tissues ultimately arises from the deformation, rearrangement, and interactions of their constituent cells. Individual cells possess measurable mechanical properties and can undergo substantial shape changes when subjected to external forces or tissue-scale stresses~\cite{kasza2007cell,Armon18}. Beyond the cellular scale, growing evidence suggests that epithelial tissues and cell collectives exhibit behaviors analogous to those of disordered materials, including glassy dynamics and jamming transitions~\cite{angelini11,park2015unjamming,bi2015density}. Epithelial tissues and cell collectives can also transition between fluid-like and solid-like deformations during important biological processes~\cite{park2015unjamming,bi2015density,mongera2018fluid}. These observations suggest conceptual parallels between living tissues and glassy materials, where deformation, flow, and solidification emerge from collective interactions among individual units. We exploit this analogy by leveraging the macroscopic deformation behavior of glass to construct physical models of tissues.

\subsection*{Glass as a Physical Modeling Platform}
Glass has been studied for millennia both as a material system and as a medium of fabrication~\cite{rasmussen2012glass}. In this study, we focus on commercially available soda-lime glass commonly used in glass art studios and hot-shop fabrication. As an amorphous solid, glass can be readily shaped, thermally driven through a well-defined transition regime, and chemically modified to tune its mechanical properties~\cite{moynihan1974dependence,scholze2012glass,berthier2011theoretical}. These characteristics enable its use in controlled forming processes, including kiln-forming techniques in which glass elements are arranged and subjected to precisely programmed heating and cooling cycles~\cite{shelby2020introduction}. In this approach, its thermal history and geometry can be independently controlled, allowing for reproducible deformation under gravitational or externally applied loads. Importantly, once the desired deformation is achieved, rapid cooling below the glass transition temperature arrests material motion, preserving the instantaneous configuration of the system. The availability of established infrastructure for glass fabrication, including glass studios, and thermal control~\cite{shelby2020introduction,dellisanti2023glass,GAS_23} makes glass a practical and experimentally accessible material for constructing physical models of epithelial tissue mechanics.

Beyond these material properties, glass also offers distinct experimental advantages: its material properties can be precisely controlled, mechanically loaded, and thermally driven through a well-defined transition regime. Moreover, standard glass-forming techniques allow rapid cooling below the glass transition temperature, effectively arresting the material and preserving snapshots of deformed configurations under load. This ability to freeze transient deformation states provides an experimentally accessible means to visualize and quantify strain localization and collective rearrangements in a physical analog system.

\subsection*{Overview of the Present Study}
Here, we fabricate glass-based physical models to study epithelial tissue mechanics under controlled loading conditions. Simplified stretch tests are performed on glass tissue analogs to simulate epithelial deformation under applied forces, and the resulting configurations are used to construct a complementary mechanics-based numerical model. To model the observed deformations, we employ a modified dynamic relaxation method, commonly used for the analysis of tensegrity structures with continuous tension elements~\cite{Ali2011}. Tensegrity structures consist of compression members (e.g., bars or struts) connected by tensioned elements (e.g., cables), with overall geometry determined by the balance of internal forces~\cite{Ali2011}. In our implementation, initial equilibrium geometries are obtained directly from cell masks extracted through image processing of the glass tissue models. These geometries are then iteratively updated in response to applied forces until new equilibrium configurations are reached. Model predictions are subsequently compared with experimentally stretched glass tissues, enabling direct quantitative comparison between physical experiments and numerical simulations.

\section*{Materials and methods}


\subsection*{Design of glass tissue models inspired by \textit{Trichoplax adhaerens}}

The morphology and deformation dynamics of \textit{Trichoplax adhaerens} inspired the design of glass-based ``tissue'' models. Prior observations of the organism’s body plan and motion\cite{prakash2021motility} informed the construction of simplified epithelial analogs. In particular, we focused on modeling the ventral epithelial layer as a confluent monolayer. Glass tissue structures were therefore composed of individual glass ``cells'' arranged to form continuous, gap-free monolayers, achieved through tack fusing.

Based on observed behaviors of \textit{T. adhaerens}~\cite{prakash2021motility}, two primary deformation modes were selected for study: lateral and radial stretching. Lateral stretching, commonly associated with asexual reproduction, involves elongation into two regions connected by a thin thread that eventually ruptures (Fig.~\ref{Fig1}). Radial stretching corresponds to deformation around internal tissue fractures. These deformation modes form the basis for the glass-based stretch experiments described below.


\subsection*{Fabrication of glass tissue monolayers}

\begin{figure}[H]
    \centering
    \includegraphics[width=1\textwidth]{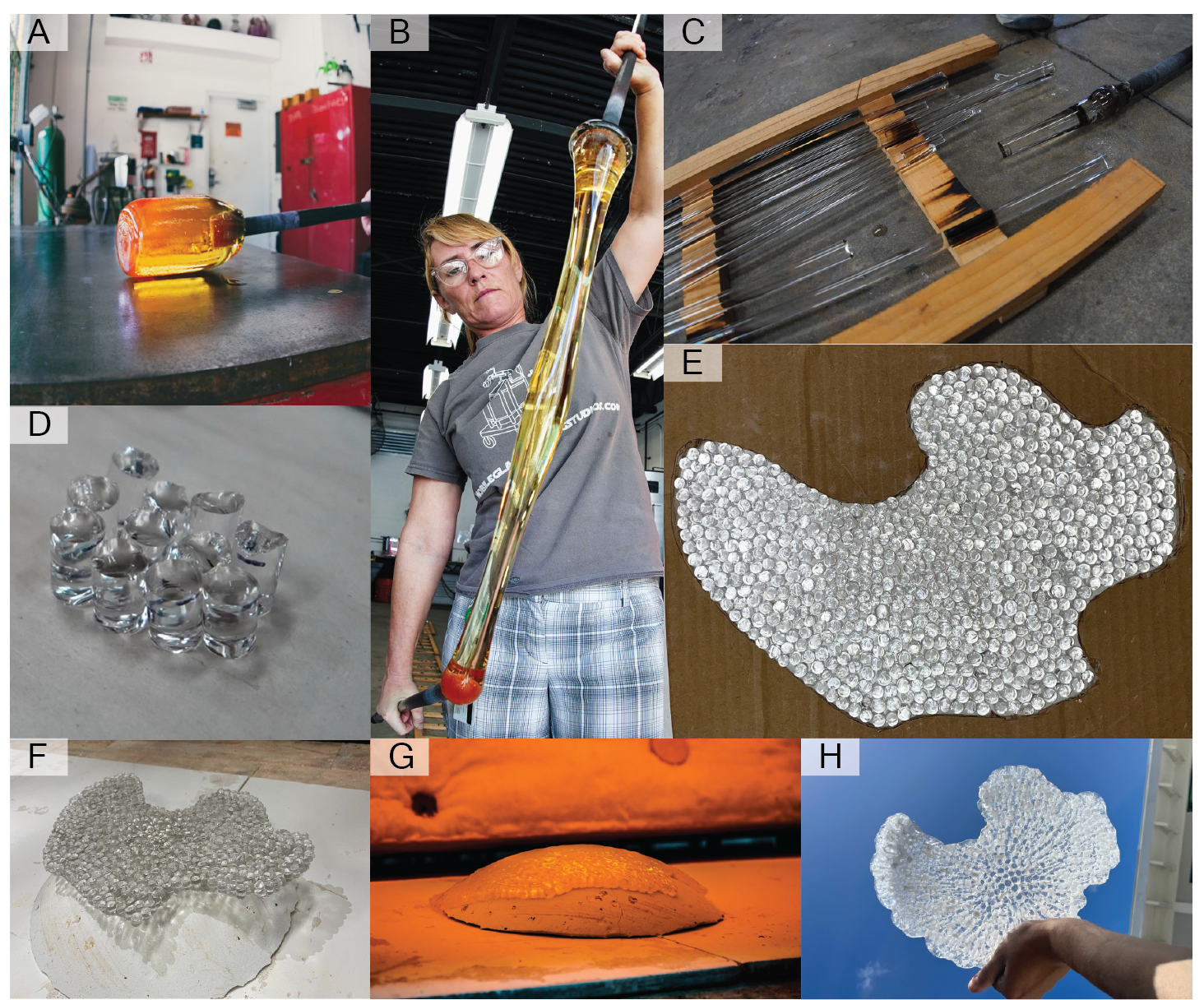}
    \caption{\textbf{Fabrication and stretching of glass ``tissue'' monolayers.} (A) Molten glass collected from the furnace on a punty and shaped into a workable mass on a marver. (B) A second pre-cooled glass post (disk) on a pontil is attached, and the glass is pulled to form a cylindrical cane. Differences in cooling between thin and thick regions enable approximately uniform diameters. (C) Pulled cane cooling on wooden supports. (D) Cane segments cut into ``cells'' and arranged into a proto-monolayer. (E) Glass cells arranged within a stencil prior to tack fusing. (F) Tack-fused monolayer positioned on a dome prior to stretching. (G) Monolayer undergoing stretching over a dome within the kiln. (H) Final stretched monolayer after cooling, preserving the deformed configuration.}
    \label{Fig 2}
\end{figure}

To replicate `cells' within a tissue, a cane-pulling technique was used. Molten glass was attached to two posts and pulled into long cylindrical rods (Fig.~\ref{Fig 2}B). These canes, up to 6~m in length and 0.5–1~cm in diameter, were placed horizontally to cool (Fig.~\ref{Fig 2}C). 

The cooled canes were then cut into segments approximately 0.95~cm in height (Fig.~\ref{Fig 2}D), which served as individual “cells” in the tissue monolayer. These glass cells were arranged in a tightly packed layer over a predefined stencil. The assembly was placed in a kiln for tack fusing, during which the glass cells were subjected to heating protocols that allow the glass cells to adhere while preserving their overall geometry. The stencil was removed once the arrangement was placed in the kiln, prior to fusion.

After cooling, the resulting glass tissue monolayer was removed from the kiln. The monolayer was placed over a curved surface substrate (plaster-silica dome) within the kiln (Fig.~\ref{Fig 2}G) to test whether the tissue will deform under gravitational loading during the glass transition temperature regime. Subsequent cooling arrested the deformation, effectively ``freezing'' the configuration and yielding stretched structures for analysis (Fig.~\ref{Fig 2}H).

\subsection*{Stretch tests}

Stretch tests were conducted on glass ``tissue'' monolayers fabricated using the kiln-forming method. To better represent cellular structure, color-overlayed glass canes were used instead of clear canes. Color overlaying involves applying a layer of colored glass over a clear glass core prior to the cane-pulling process. After pulling and cutting, the resulting glass “cells” consist of a colored boundary representing the cell membrane and a clear core representing the cell interior (Fig.~\ref{Fig 3}A and Fig.~\ref{Fig 4}A). These glass “cells” were then arranged and tack fused into monolayers with geometries tailored to the type of stretch test performed. 

The analysis involved quantifying the geometry of the glass “cells” in the pre and post deformed monolayers. Samples were imaged on a light table with a calibrated length scale for reference. The acquired images were processed in MATLAB to generate binary masks of individual cells, aided by the color-overlayed canes for clear boundary identification. From these masks, cell area and eccentricity were extracted. Measurements from multiple cells were compiled and analyzed, and variability across replicates was quantified using standard deviation. Appropriate statistical tests were applied for each type of stretch test based on the nature of the data. Statistical significance was denoted as follows: *** for $p < 0.001$, ** for $p < 0.01$, * for $p < 0.05$, † for $p < 0.10$ (marginal significance), and n.s. for non-significant results ($p \geq 0.10$).

\subsubsection*{Lateral stretch}

\begin{figure}[H]
    \centering
    \includegraphics[width=0.95\textwidth]{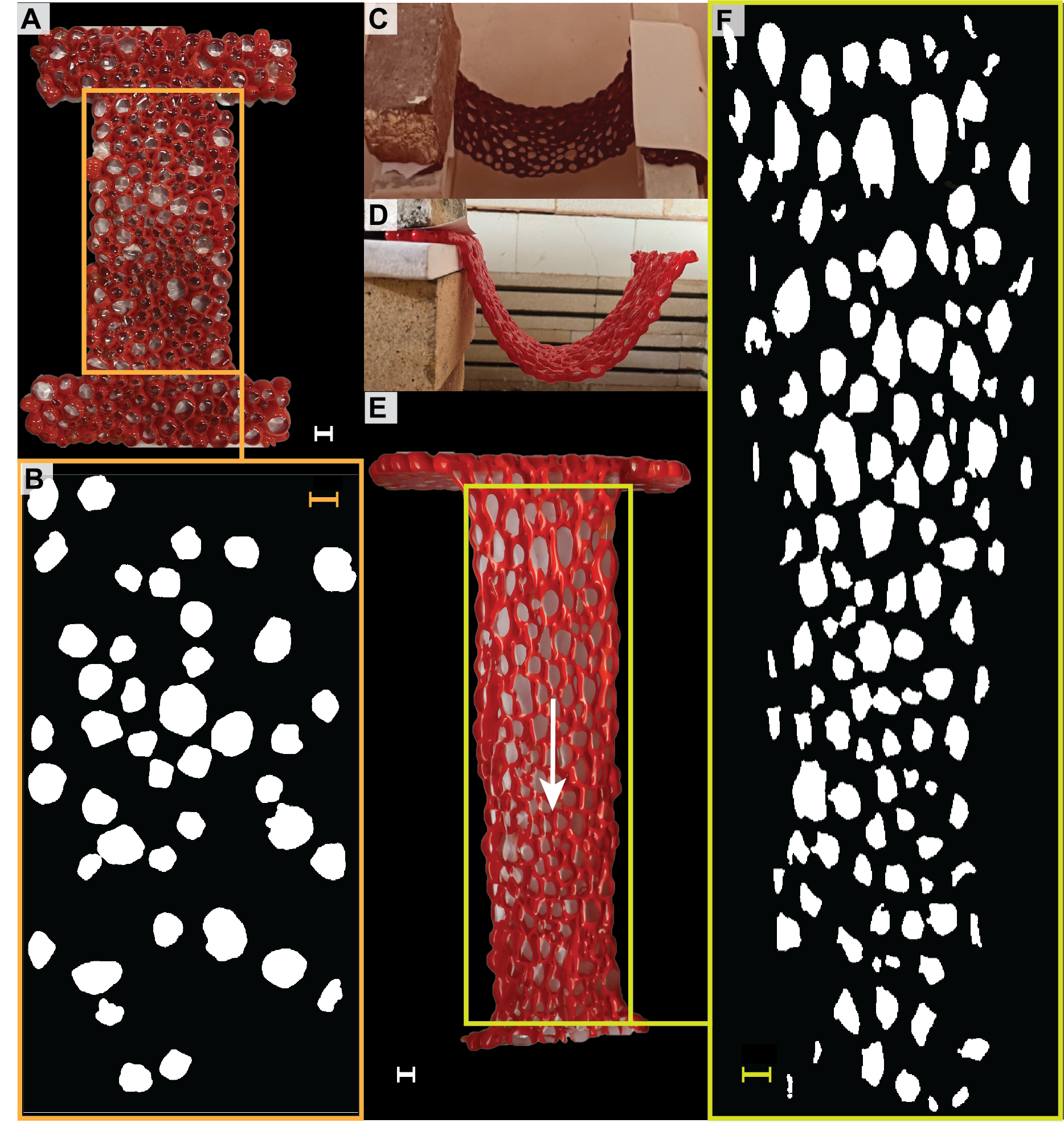}
    \caption{{\textbf{Lateral stretching and image analysis of glass “tissue” monolayers.}} (A) Tack-fused glass “tissue” monolayer. (B) Binary mask obtained from image processing of the monolayer (zoomed-in view). (C) Monolayer elevated on a support to induce sagging deformation. (D) Monolayer after sag-induced stretching, prepared for subsequent straightening. (E) Final straightened and stretched monolayer. (F) Binary mask obtained from image processing of the stretched monolayer (zoomed-in view). All scale bars correspond to 1 cm.}
    \label{Fig 3}
\end{figure}

Lateral stretch tests were designed to investigate ductile deformation in the glass “tissue” monolayers. Inspired by fission events in \emph{T. adhaerens}, lateral stretch geometries were fabricated using a capital ``I''-shaped stencil. The monolayer consisted of a central rectangular region (7~cm $\times$ 21~cm; area = 147~cm\textsuperscript{2}) connected to four 3~cm $\times$ 3~cm square “tags” that provided additional surface area for suspension (Fig.~\ref{Fig 3}).

Lateral stretching was applied through a two-step heating process in the kiln. The tack-fused monolayer was supported at the ends, allowing the central region to sag under gravity as the temperature approached the glass transition regime. During this sagging phase, separation of one of the tag regions from the central rectangle was occasionally observed. Following the initial sag-induced deformation, the monolayer was reheated while fixed at one end, allowing further stretching and partial straightening. After cooling, the deformed monolayer was retrieved for subsequent analysis (Fig.~\ref{Fig 3}). Cell area and eccentricity following lateral stretching were quantified and visualized using violin plots (Fig. \ref{Fig 5} A, B). Statistical differences between conditions were assessed using a two-sided Wilcoxon rank-sum test.

\subsubsection*{Radial--uniaxial stretch}

\begin{figure}[H]
    \centering
    \includegraphics[width=1\textwidth]{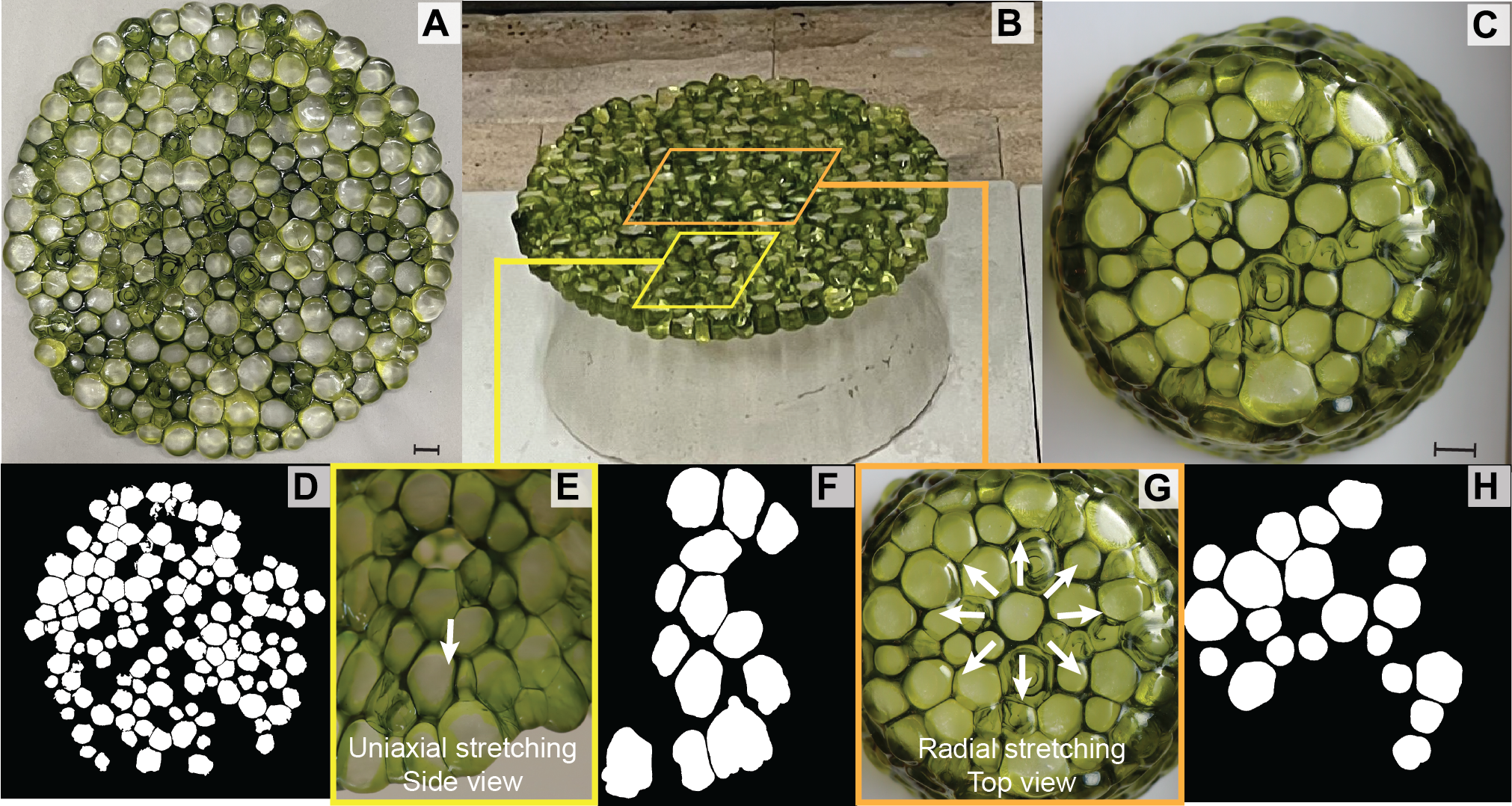}
    \caption{\textbf{Radial–uniaxial deformation and image analysis of glass “tissue” monolayers.} (A) Disk-shaped tack-fused monolayer prior to deformation. (B) Monolayer positioned on a dome for stretching. (C) Monolayer after deformation. (D) Binary mask obtained from image processing of the undeformed monolayer (A). (E) Region of the monolayer exhibiting uniaxial stretching. (F) Corresponding binary mask obtained from image processing of (E). (G) Region of the monolayer exhibiting radial stretching. (H) Corresponding binary mask obtained from image processing of (G). All scale bars correspond to 1 cm.}
    \label{Fig 4}
\end{figure}

Circular glass “tissue” monolayers for radial–uniaxial stretching were fabricated by assembling individual glass “cells” into circular geometries. Circular monolayers were selected to minimize geometric anisotropy. The resulting tissues had an area of approximately 250~cm\textsuperscript{2} and a diameter of 17.8~cm (Fig. \ref{Fig 4}). Initial plaster–silica dome prototypes were replaced with conical frustums to enable controlled gravitational loading. Frustums were designed in Rhinoceros 3D and fabricated via CNC carving~\cite{rhino3d}. Negative molds were produced, sanded, and sealed prior to casting. A plaster–silica mixture (1:1:1 by weight with water) was poured into the molds, mechanically agitated to minimize air entrapment, and allowed to cure fully before demolding. The resulting frustums were subsequently dried and used in kiln-based deformation experiments~\cite{dellisanti2023glass}.

Kiln-based stretching was performed under a controlled thermal protocol (Table \ref{Table1}). Glass monolayers were centered on the frustums and heated to the transition regime, allowing deformation under gravitational loading. The system was first heated beyond the transition temperature to 566~\textdegree C over approximately 12~h to ensure thermal equilibration and complete drying of the molds (Oceanside Studio Glass Protocol)~\cite{moynihan1974dependence,dellisanti2023glass} . The temperature was then increased to 871~\textdegree C, at which point the glass softened and conformed to the frustum geometry. Heating was terminated once the desired deformation was achieved, and the kiln was transitioned to an annealing cycle. Controlled cooling over approximately 7~h arrested the deformation, preserving the final configuration for subsequent analysis. Data distributions of cell area and eccentricity (Fig. \ref{Fig 5} C, D) were visualized using violin plots. Statistical differences between groups were assessed using the Kruskal–Wallis test, followed by Dunnett’s multiple comparisons test (vs control).


\begin{table}[h]
\centering
\caption{Experimental parameters for glass tissue stretch tests (Oceanside Studio, Olympic Color Rods) ~\cite{dellisanti2023glass}}
\label{Table1}
\begin{tabular}{|p{3cm}|p{3cm}|p{3cm}|p{3cm}|}
\hline
\textbf{Experiment Type} & \textbf{Glass Colors} & \textbf{Coefficient of Expansion ($10^{-7}$/~\textdegree C)} & \textbf{Time at 760~\textdegree C} \\
\hline
Lateral Stretch & Red, Clear & 96, 95 & 45 min + 30 min \\
\hline
Radial Stretch & Red, Green, Clear & 96, 96, 95 & 35 min \\
\hline
\end{tabular}
\end{table}

\subsubsection*{Mathematical modeling of glass “tissue” monolayers}

\begin{figure}
    \centering
\includegraphics[width=0.5\textwidth]{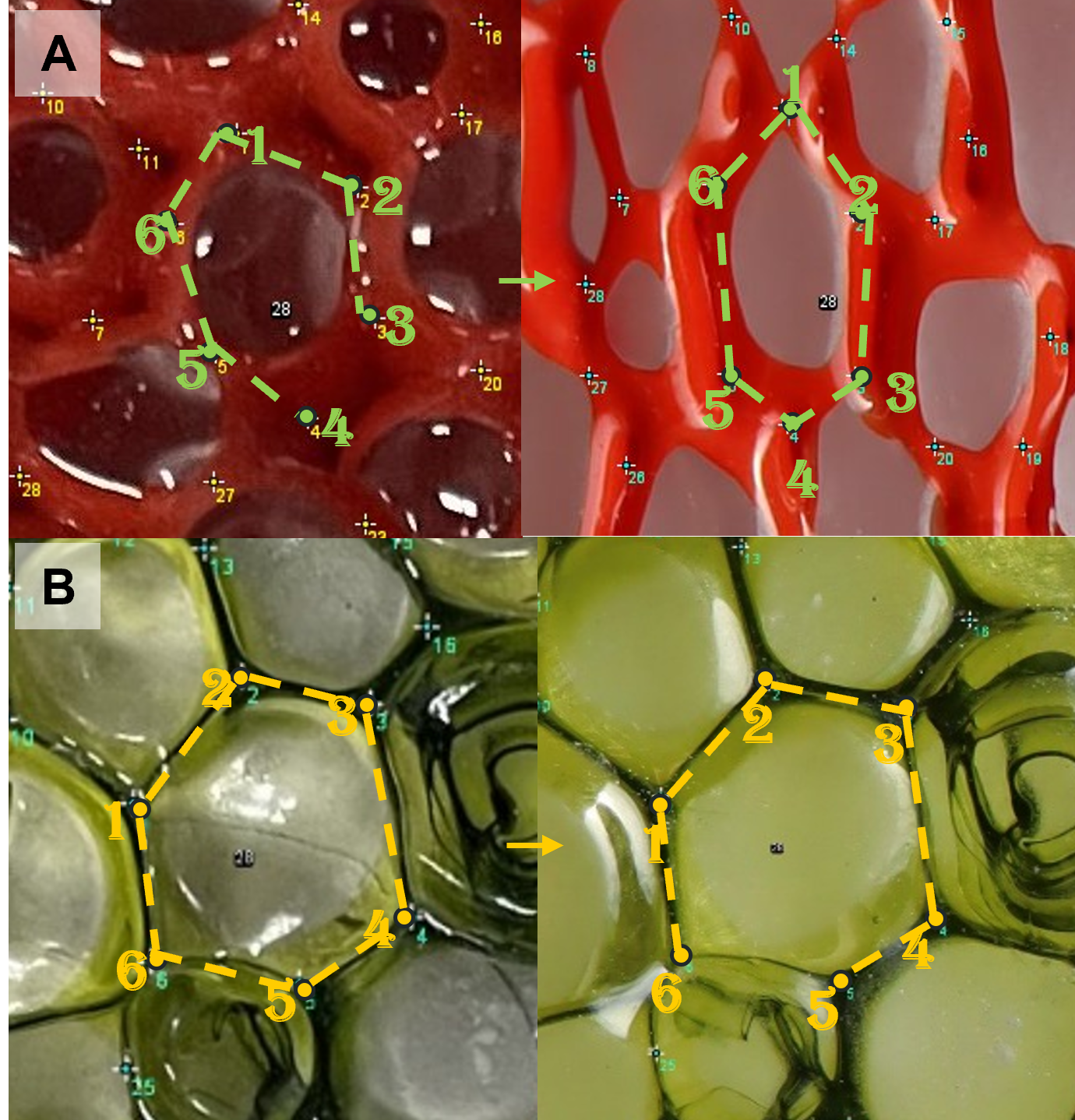}
    \caption{\textbf{Experimental cell geometries used as inputs for the tensegrity model.} (A) Representative cell from a glass “tissue” monolayer before (left) and after (right) lateral stretching. (B) Representative cell from a glass “tissue” monolayer before (left) and after (right) radial stretching. Dashed polygons connect corresponding cell vertices identified through image analysis. The measured vertex coordinates define the network geometry used in the tensegrity-based dynamic relaxation model and provide a direct link between experimental cell deformations and numerical simulations.}
    \label{Fig 5}
\end{figure}

To complement the experimental measurements, a mathematical physics-based model was developed to quantify deformations in the cellular geometry of the glass “tissue” monolayers. The model is based on a tensegrity-inspired framework representation~\cite{aloui2018generation}. In this framework, glass “tissue” monolayers are modeled as a network of hexagonal cell units described by a set of nodes, and their behavior is modeled through effective connections that also capture the mechanical coupling between neighboring cells (Fig.~\ref{Fig 5}). The axial stiffness of the connections was established by incorporating the glass Young's modulus (72 GPa)~\cite{scholze2012glass,rasmussen2012glass} and the cross-sectional area, the latter of which was derived from the thickness of the glass tissue. The resulting network encodes the geometry of the tissue and allows deformation under applied loading to be computed.

To model the observed deformations, we used the dynamic relaxation method, a numerical approach commonly used for the form-finding and analysis of tensile structures~\cite{Ali2011}. The method transforms a static equilibrium problem into a pseudo-dynamic one. The solution is obtained by tracing nodal motion until the structure reaches a stable static equilibrium via damping. The governing equation is: 
\begin{equation}
        {\mathbf{F}_{ext}} - {\mathbf{F}_{int}} = M\mathbf{a} + C \mathbf{v} 
    \end{equation}
where ${\mathbf{F}_{ext}}$ and ${\mathbf{F}_{int}}$ are the external and internal nodal loads. Here, $\mathbf{a}$ and $\mathbf{v}$ represent acceleration and velocity, while $M$ and $C$ are fictitious mass and damping parameters optimized for numerical stability and convergence~\cite{papadrakakis1981method}. The final configuration of the network is in equilibrium with the external loads. External loading conditions corresponding to uniaxial and radial stretching were applied to the glass cell network, and the resulting deformed configurations were computed. Model outputs were analyzed using the same metrics as the experiments, including cell area and eccentricity, enabling direct quantitative comparison between experimental measurements and simulations. Simulations were conducted using MATLAB, building upon a robust methodology previously explored across various systems including dielectric elastomers, and tensegrity structures~\cite{feron2023experimental,aloui2019theoretical,segal2015multi,siu2013dynamic}.

\section*{Results}

\subsection*{Spatial variation in deformation modes}

Distinct deformation patterns were observed under different stretching conditions. In lateral stretching (Fig.~\ref{Fig 3}), deformation is relatively uniform across the monolayer, with cells undergoing elongation in a consistent direction. In contrast, radial–uniaxial loading (Fig.~\ref{Fig 4}) produces pronounced spatial variation within the same monolayer. Regions near the center exhibit predominantly radial stretching, characterized by approximately isotropic expansion of cell geometry, while peripheral regions show uniaxial stretching, with cells elongating preferentially along a single direction.

This contrast highlights the role of geometry and boundary conditions in determining the spatial distribution of deformation, with radial–uniaxial loading giving rise to heterogeneous cell-scale responses within a single tissue.

\subsection*{Cell-scale deformation under lateral and radial stretching}

Glass “tissue” monolayers were subjected to lateral and radial–uniaxial stretching, and the resulting cell-scale deformations were quantified. Binary masks generated from image processing of the monolayers (Figs.~\ref{Fig 3} and \ref{Fig 4}) enabled measurement of individual ``cell'' area and eccentricity before and after deformation.

\begin{figure}[H]
    \centering
    \includegraphics[width=1\textwidth]{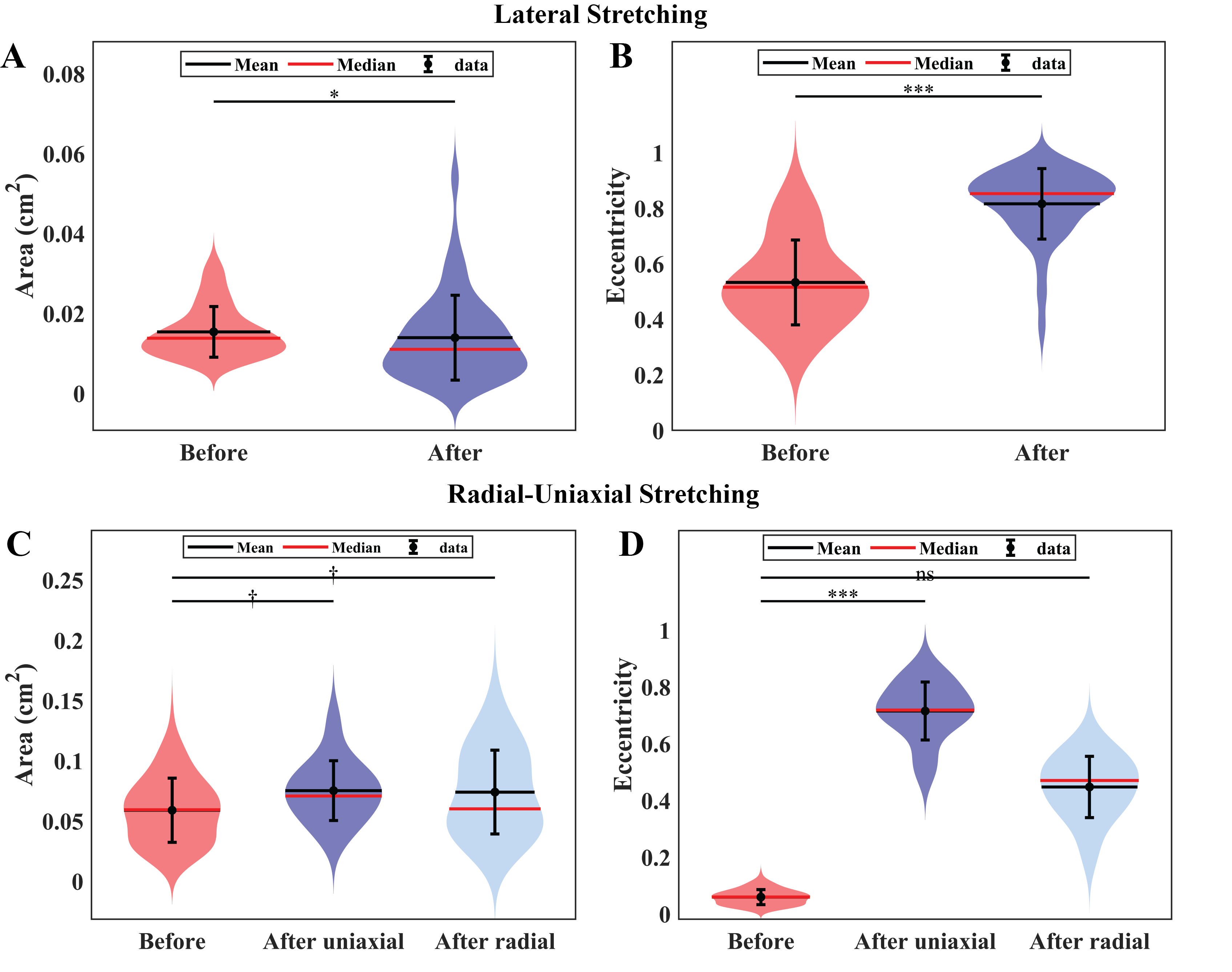}
    \caption{\textbf{Quantification of cell-scale deformation in glass “tissue” monolayers under different stretching conditions.} Violin plots showing (A) cell area before and after lateral stretch, (B) cell eccentricity before and after lateral stretch, (C) cell area before and after radial–uniaxial stretch, and (D) cell eccentricity before and after radial–uniaxial stretch. Statistical comparisons between groups were performed using two-sided non-parametric tests (Wilcoxon rank-sum test for pairwise comparisons and Kruskal–Wallis test for multi-group comparisons; sample sizes vary across metrics). Cell area and eccentricity under lateral stretch showed significant changes (p = 0.027 and p $<$ 0.0001, respectively). For radial–uniaxial stretch, Dunnett’s multiple comparisons test (vs Before) was used for post hoc comparisons following Kruskal–Wallis testing. Cell area showed a marginally significant difference for Uniaxial vs Before (p = 0.088) and Radial vs Before (p = 0.050). Cell eccentricity showed significant and non-significant differences for Uniaxial vs Before (p $<$ 0.0001) and Radial vs Before (p = 0.124), respectively.}
    \label{Fig 6}
\end{figure}

Both cell area and eccentricity were quantified from the stretching experiments (Fig.~\ref{Fig 6}). Cell eccentricity generally increased following deformation across all stretching conditions (Fig.~\ref{Fig 6}B,D). Increases in eccentricity were statistically significant for lateral stretching and radial-uniaxial stretching, whereas radial stretching alone did not produce a statistically significant increase.

In lateral stretching (Fig.~\ref{Fig 6}A,B), cells exhibited modest but statistically significant changes in area, but with moderate increases in eccentricity, consistent with anisotropic deformation. The broader eccentricity distributions observed after lateral stretching suggest greater variability in the deformation response among cells. However, in radial–uniaxial stretching (Fig.~\ref{Fig 6}C,D), while cells exhibited measurable changes in area, they showed much larger increases in eccentricity, reflecting stronger deformation associated with radial expansion and localized uniaxial stretching. Radial stretching alone produced substantially smaller changes in eccentricity than uniaxial stretching. The eccentricity distributions under radial–uniaxial loading are narrower, indicating decreased heterogeneity. The magnitude of the observed changes depended not only on the deformation mode, but also on the spatial location of cells within the monolayer.

\subsection*{Model–experiment comparison}

\begin{figure}[H]
    \centering
    \includegraphics[width=1\textwidth]{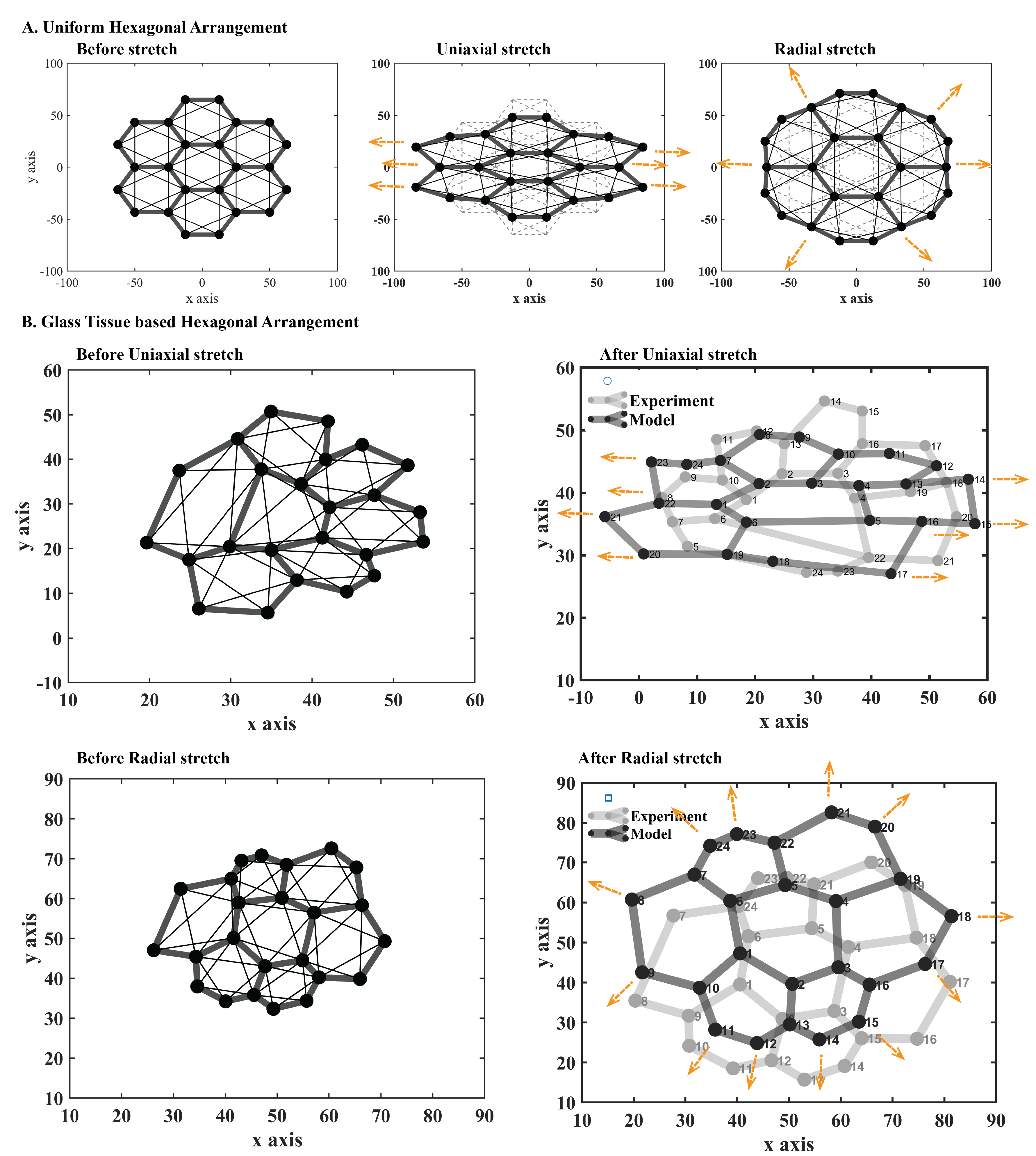}
    \caption{\textbf{Comparison between model predictions and experimental deformation of glass “tissue” monolayers.} (A) Idealized hexagonal lattice subjected to uniaxial and radial stretching in the model, showing the imposed deformation modes. (B) Top Row: Glass “tissue” network before and after uniaxial stretching, with experimental node positions overlaid with corresponding model predictions. Bottom Row: Glass “tissue” network before and after radial stretching (with an offset for visibility), with experimental node positions overlaid with corresponding model predictions. In both deformation modes, the model captures the qualitative changes in network geometry observed in the experiments, including anisotropic elongation under uniaxial loading and isotropic expansion under radial loading.}
    \label{Fig 7}
\end{figure}

To interpret the observed deformation patterns, a physics-based tensegrity model was developed and used to simulate lateral and radial stretching. Initial investigations utilized an idealized network composed of uniform hexagonal cells to isolate the effects of loading geometry from structural heterogeneity (Fig.~\ref{Fig 7}A).

Application of uniaxial loading produced elongation along the loading direction accompanied by contraction in the transverse direction, whereas radial loading generated a more isotropic outward expansion of the network (Fig.~\ref{Fig 7}A). In both cases, the model reproduced the expected deformation response while preserving network connectivity, providing a baseline for interpreting the more heterogeneous tissue architectures examined experimentally.

To investigate the role of structural heterogeneity present in the experiments, simulations were performed on networks reconstructed from image-processed glass ``tissue'' monolayers (Fig.~\ref{Fig 7}B). In contrast to the idealized hexagonal lattice, these networks contain realistic variations in cell geometry and connectivity. Under uniaxial loading, the model predicted a non-uniform deformation field characterized by differential node displacements throughout the network (Fig.~\ref{Fig 7}B, Top Row). Comparison of the simulated and experimental configurations showed agreement in the overall direction of deformation, although differences in local displacement magnitudes and network geometry were apparent. These results suggest that the tensegrity framework captures the primary mechanical response of the glass tissue during uniaxial stretching while leaving room for further refinement.

A high degree of qualitative agreement was observed between the simulated and experimental configurations under radial loading (Fig.~\ref{Fig 7}B, Bottom Row).  The model predicted coordinated outward motion of both peripheral and interior nodes, producing an overall expansion of the network. The resulting simulated geometry closely matched the experimentally observed configuration, with many nodes exhibiting comparable displacement directions and relative magnitudes. Together, these results demonstrate that the tensegrity framework captures the principal deformation characteristics of glass tissue monolayers under both uniaxial and radial loading conditions.

\begin{figure}[H]
    \centering
    \includegraphics[width=1\textwidth]{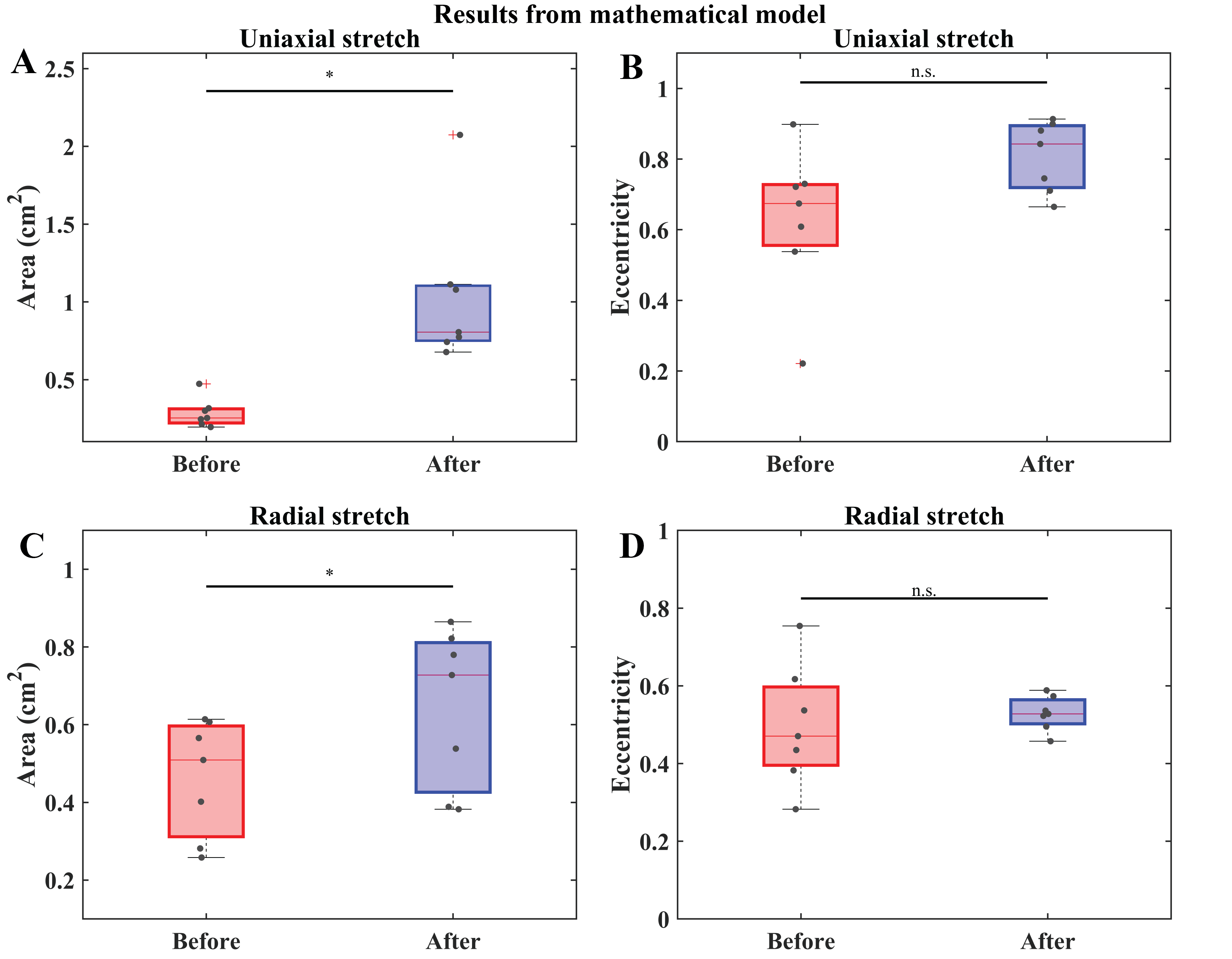}
   \caption{\textbf{Model-based quantification of cell-scale deformation under uniaxial and radial stretching.} Box plots show the distribution of cell area (A,C) and eccentricity (B,D) before and after uniaxial (A,B) and radial (C,D) stretching for n = 7 paired samples. Statistical significance was assessed using a two-sided Wilcoxon signed-rank test (n = 7). Area and eccentricity under uniaxial stretch showed a significant (*, p = 0.016) change and a non-significant (n.s., p = 0.109) change respectively.  Area and eccentricity under radial stretch showed significant (*, p = 0.016) change and non-significant (n.s., p = 0.578) change respectively.}
    \label{Fig 8}
\end{figure}

Quantitative analysis of the model predictions further supports agreement between simulations and experiments (Fig.~\ref{Fig 8}). Under both uniaxial and radial loading, the model predicts significant increases in cell area following deformation (Wilcoxon signed-rank test, $p = 0.016$ for both loading modes). In contrast, cell eccentricity exhibited comparatively small and statistically non-significant changes under either loading condition. Although the model underestimates the eccentricity changes observed experimentally, it reproduces the overall increase in cell area and captures the dominant geometric features of the deformation response.

Although variability is present due to differences in local network geometry and boundary conditions, the model successfully reproduces the principal statistical trends observed in the glass monolayers. Together with the qualitative agreement shown in Fig.~\ref{Fig 7}, these results demonstrate that the tensegrity framework captures the principal network-scale deformation patterns observed in the experiments. The remaining differences, particularly in cell eccentricity, likely reflect simplifications in the current model and provide opportunities for future refinement.

\section*{Discussion}

\subsection*{Key Findings}
This study demonstrates that glass-based “tissue” monolayers provide a controllable physical medium for investigating cell-scale deformation in epithelial systems. By combining controlled fabrication, thermal loading, and quantitative image analysis, we show that glass tissues undergo measurable changes in cell geometry, including area and eccentricity, under lateral and radial–uniaxial stretching.

These deformation modes exhibit distinct spatial patterns, with relatively uniform elongation under lateral loading and heterogeneous, position-dependent deformation under radial–uniaxial loading. A tensegrity-inspired physical model reproduces the major qualitative deformation patterns and several quantitative features of the experiments, providing a useful framework for interpreting the observed behavior. While the model successfully captures changes in cell area and overall network geometry, it underestimates the eccentricity changes observed experimentally.

A key result of this work is the emergence of distinct deformation regimes that depend on geometry and boundary conditions. Under lateral stretching, deformation is largely uniaxial, leading to elongation of cells along a preferred direction. In contrast, radial–uniaxial loading produces spatially varying deformation, with isotropic expansion near the center and anisotropic elongation in peripheral regions. These observations highlight how global geometry and loading conditions give rise to heterogeneous cell-scale responses within a continuous tissue, a feature that is difficult to isolate in living systems.

\subsection*{Glass-Based Physical Models for Tissue Mechanics}

The use of glass as a physical analog is motivated by parallels between cellular collectives and materials that exhibit fluid-to-solid transitions. Tissue-scale deformation ultimately arises from the mechanical behavior, interactions, and rearrangements of individual cells. At larger scales, epithelial tissues have been shown to exhibit phenomena associated with disordered matter, including glassy dynamics, jamming transitions, and fluid-to-solid state changes~\cite{angelini11,park2015unjamming,bi2015density,mongera2018fluid}. Glass-forming materials similarly undergo transitions between fluid-like and solid-like behavior while maintaining structural integrity. In our experiments, deformation of the glass monolayers under gravity produces coordinated changes in cell geometry and network structure that are qualitatively analogous to collective mechanical responses observed in epithelial tissues. While the analogy is not exact and does not capture active biological processes, these similarities suggest that glass-based systems can provide useful physical models for investigating tissue-scale deformation and mechanical behavior.

An important advantage of this approach is the ability to arrest deformation through controlled cooling, effectively “freezing” the system at defined stages of strain. This enables direct visualization and quantification of intermediate deformation states, which are often difficult to access in living tissues. The glass platform therefore complements existing experimental and computational approaches by providing a macroscopic, tunable system in which geometry, loading, and material properties can be systematically varied.

Although the model does not fully reproduce all aspects of the experimental response, particularly the magnitude of eccentricity changes, it successfully captures the dominant deformation modes and overall network-scale geometry. This agreement suggests that geometric constraints and connectivity alone can account for the observed deformation patterns, even in the absence of active biological processes.

\subsection*{Limitations and Future work}
Despite these advantages, several limitations should be noted. The glass system is a passive material and does not capture active processes such as cellular force generation, adhesion dynamics, or biochemical signaling. In addition, the current model simplifies tissue architecture and does not account for heterogeneity in cell properties, dynamic remodeling, or active cellular processes. These simplifications likely contribute to the differences between model predictions and experimental measurements, particularly for cell eccentricity. Future work could incorporate more complex geometries, time-dependent loading, and coupling with active or fluid components to better approximate biological conditions. It could also incorporate \textit{in situ} imaging and continuous monitoring during deformation, enabling reconstruction of the full deformation trajectory rather than only the initial and final arrested states.

\section*{Conclusion}

Glass “tissue” monolayers developed in this study provide a controllable experimental platform for modeling cell-scale deformation in epithelial systems. By constructing cohesive monolayers and quantifying changes in cell geometry under thermal loading, tissue deformation can be directly visualized and measured. The ability to precisely control experimental conditions and to arrest deformation through rapid cooling enables access to intermediate deformation states that are difficult to capture in living tissues.

These results establish glass-based systems as promising physical analogs for studying tissue mechanics. Beyond its relevance to tissue mechanics, this work highlights a novel use of glass fabrication as a quantitative experimental platform for studying biological deformation. By repurposing techniques traditionally associated with glass art and studio practice, the glass shop becomes a laboratory for investigating mechanical phenomena in living systems. More broadly, the use of the glass shop as a laboratory for investigating biological phenomena highlights a productive intersection between art, materials science, and biophysics, with opportunities for interdisciplinary research and scientific communication~\cite{GAS_23,PhyToday,madhu2024APS}.

\section*{CRediT authorship contribution statement}

Gopika Madhu: Methodology, Investigation, Formal analysis, Data curation, Visualization, Writing – original draft, Writing – review \& editing. 

Carolyn Delli-Santi: Conceptualization, Methodology, Investigation, Data curation, Visualization, Writing – original draft, Writing – review \& editing.

Jenna Efrein: Conceptualization, Methodology, Investigation, Resources, Project administration, Funding acquisition, Supervision, Writing – review \& editing. 

Prannoy Suraneni: Methodology, Resources, Funding acquisition, Supervision, Writing – review \& editing. 

Landolf Rhode-Barbarigos: Methodology, Investigation, Formal analysis, Data curation, Supervision, Writing – review \& editing.

Vivek N. Prakash: Conceptualization, Methodology, Resources, Project administration, Funding acquisition, Supervision, Writing – review \& editing.

\section*{Acknowledgments}

The authors thank Leah Henseler and Katie Hastings for assistance in the glass shop, and members of the Prakash Lab for helpful discussions. J.~E., P.~S., and V.~N.~P. acknowledge funding support from the University of Miami Laboratory for Integrative Knowledge (U-LINK) project “Engineering Corals for Climate Change Resilience” and the University of Miami Provost’s Award for Collaborative Teaching. V.~N.~P. also acknowledges start-up funding support from the University of Miami.





\bibliography{reference}


\section*{Supporting information}


\setcounter{figure}{0} 
\renewcommand{\thefigure}{S\arabic{figure}}

\setcounter{table}{0} 
\renewcommand{\thetable}{S\arabic{table}}

All colored glass used in this study was purchased from Olympic Color Rods. Reichenbach Fire Red (SKU: R-147) was selected to model more brittle behavior, while Reichenbach Olive (SKU: R-023) was used to represent more compliant, elastic behavior. Clear glass was supplied by Oceanside Studio Glass Nuggets, with a coefficient of thermal expansion of $95 \times 10^{-7}/$°C over the temperature range 0–300°C \cite{dellisanti2023glass}.

To replicate individual cells, a color-overlayed cane pulling technique was employed. In this method, glass canes, long cylindrical rods with diameters of approximately 0.5–1~cm, are fabricated with a clear core and a colored outer layer. The process involves gathering molten clear glass from a furnace onto a steel rod (punty) to form a cylindrical mass, which is subsequently overlayed with a mass of colored glass. The combined glass mass is repeatedly heated and shaped to ensure uniform distribution of the color layer. Controlled cooling between heating steps increases the viscosity of the glass, allowing the formation of a stable, elongated geometry suitable for subsequent stretching into cane.

\paragraph*{Hotshop Fabrication Details\\}

A pre-heated segment of colored glass (held at approximately 519~°C) was retrieved from a kiln and brought to working temperature before being applied to the surface of the clear glass preform. The colored glass was then distributed uniformly along the length of the preform through controlled heating and shaping. This process involved alternating reheating with mechanical shaping on a steel surface (marver) to ensure even coating and to promote thermal equilibration throughout the glass mass.

A secondary glass support (“post”) was prepared by gathering molten glass from the furnace and shaping it into a widened form. The tip of this form was pressed against a steel surface (marver) to create a flat circular interface for attachment to the opposite end of the glass preform. The marver acts as a heat sink, rapidly extracting heat and increasing the local stiffness of the glass, thereby providing a stable support during the subsequent stretching process.

The glass preform was then attached to the prepared support and elongated to initiate cane formation. Initial stretching was assisted by gravity, followed by controlled, symmetric pulling from both ends to extend the glass into a long, cylindrical rod. As the glass was drawn out, its diameter decreased while maintaining a continuous core–shell structure. This process yielded canes up to several meters in length before the glass cooled sufficiently to become rigid.

Upon completion of the stretching process, the glass cane was transferred to wooden supports to cool under controlled conditions. These supports act as thermal buffers, reducing the cooling rate and minimizing the development of residual stresses. Once partially cooled and sufficiently rigid, the cane was detached from the supporting rods by locally snapping the glass using steel tweezers. The cane was then segmented into shorter lengths suitable for handling, with sections of approximately 25~cm produced for subsequent processing. The resulting segments exhibited diameters of 0.5–1~cm, with a clear core and a thin, uniformly distributed colored outer layer. The segments were then allowed to cool completely without further disturbance to preserve structural integrity.

After cooling, the cane segments were further processed using a mechanical glass chopper (Astral Glass Studio, LLC) to produce uniformly sized elements. The chopper, equipped with an adjustable backstop, ensured consistent segment length during cutting. The glass cane was positioned against the backstop and cut between two circular blades, producing segments of approximately 0.95~cm in height through controlled snapping. This process was repeated until all cane was sectioned into cylindrical “cells” with consistent height and diameter.

Glass “cells” were arranged upright in close-packed configurations using cardboard stencils to form monolayer assemblies. The stencils were placed on kiln shelves to facilitate handling and loading into a chest kiln. Prior to assembly, the kiln shelves were coated with boron nitride aerosol to prevent adhesion and allowed to dry completely. Once the stencils were fully populated, they were carefully removed to preserve the arrangement, and the assembled monolayers were transferred into the kiln for tack fusing. The chest kilns used in this study consist of insulated enclosures with heating elements embedded in the side walls and hinged lids, allowing uniform thermal treatment. Multiple monolayers (up to eight) could be processed simultaneously in a single firing cycle.

\begin{figure}[H]
    \centering
    \includegraphics[width=1\textwidth]{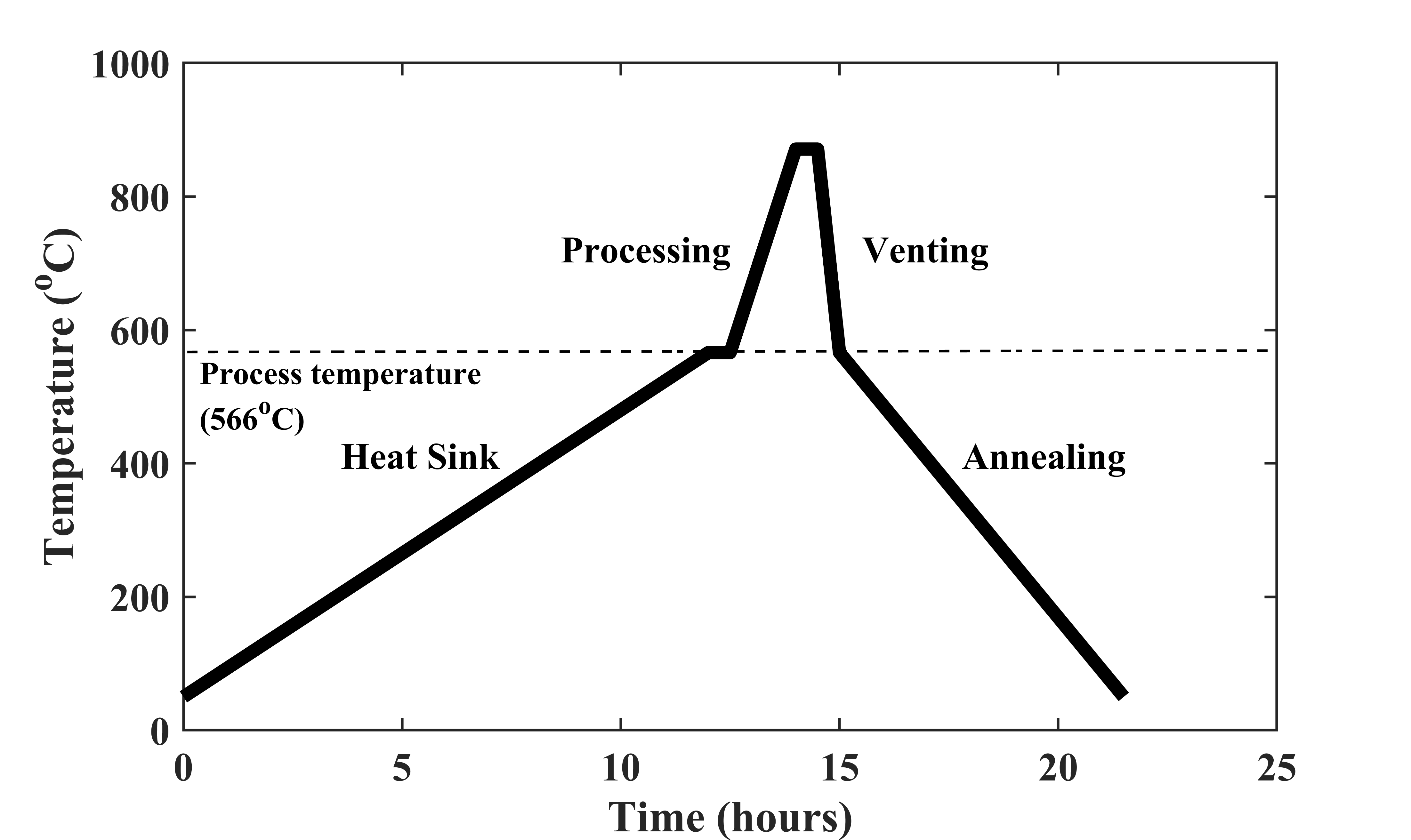}
    \caption{\textbf{Thermal program used for tack fusing and deformation of glass “tissue” monolayers.} The kiln temperature is ramped from room temperature to the glass transition regime, followed by a short dwell period to enable softening and interfacial bonding between adjacent glass elements. The system is then rapidly vented to arrest deformation and subsequently cooled under controlled conditions through an annealing phase to relieve residual stresses.}
    \label{fig:Figures/Final figures/FigureS1}
\end{figure}

Tack fusing is a kiln-forming process used to bond adjacent glass elements while preserving their overall geometry, with only minor rounding of sharp edges. The thermal protocol (Fig.~\ref{fig:Figures/Final figures/FigureS1}) begins with a slow temperature ramp to approximately 566~°C (1050~°F) over $\sim$12~hours, allowing thermal equilibration and the development of a heat sink within the kiln’s refractory materials. The temperature is then increased more rapidly to 760~°C (1400~°F), exceeding the strain point of the glass (518~°C), to enable interfacial bonding between neighboring elements and formation of a cohesive monolayer.

At the peak temperature, the glass was monitored to determine the completion of tack fusing, which typically occurred within 20–30~minutes as edges softened and bonded. Slight variations in fusion time were observed between compositions, with the stiffer red glass requiring marginally longer dwell times than the more compliant green glass (by approximately 5~minutes). Once bonding was achieved, the kiln was manually vented to rapidly reduce the temperature from the working range ($\sim$760~°C) to just above the annealing temperature (approximately 518~°C). This rapid temperature drop arrests further deformation while maintaining the structural integrity of the fused monolayer.

Following venting, the system was cooled gradually over $\sim$7~hours through an annealing schedule to relieve residual stresses and prevent cracking. After complete cooling, the resulting glass monolayers were removed from the kiln and inspected to ensure uniformity and the absence of defects prior to subsequent deformation experiments.

\begin{figure}[h]
    \centering
    \includegraphics[width=1\textwidth]{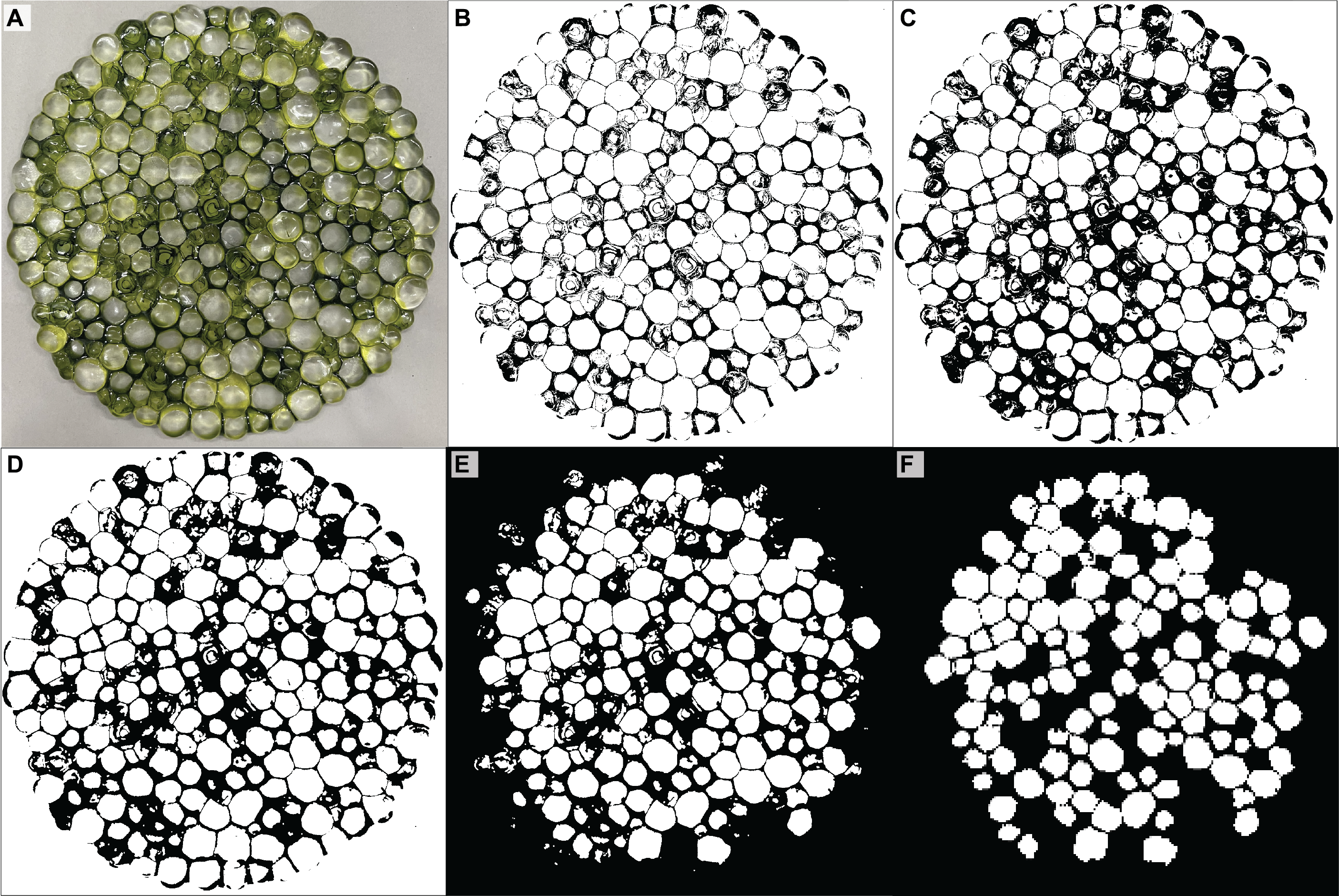}
    \caption{\textbf{Image processing workflow used to generate binary masks from glass “tissue” monolayers.} (A) Original image of the glass epithelium. (B) Adaptive binarization. (C) Active contour refinement. (D) Mask thresholding. (E) Removal of boundary artifacts and hole filling. (F) Area-based filtering of segmented regions to obtain the final binary mask used for quantitative analysis.}
    \label{fig:Figures/Final figures/FigureS2.png}
\end{figure}

\paragraph*{Image Processing\\}
\label{S1}
Image processing was performed using the Image Processing Toolbox in MATLAB. Raw images of the glass “tissue” monolayers were first converted to grayscale and subjected to adaptive binarization to distinguish cell interiors from boundaries (Fig.~\ref{fig:Figures/Final figures/FigureS2.png}A,B). The resulting masks were refined using active contour methods to improve segmentation of individual cells (Fig.~\ref{fig:Figures/Final figures/FigureS2.png}C). Additional thresholding and morphological operations were applied to remove boundary artifacts and fill holes within segmented regions (Fig.~\ref{fig:Figures/Final figures/FigureS2.png}D,E). Finally, small or irregular regions were filtered based on area to obtain clean binary masks representing individual cells, which were used for subsequent quantitative analysis of cell area and eccentricity (Fig.~\ref{fig:Figures/Final figures/FigureS2.png}F).

\paragraph*{Validation\\}

To validate the accuracy of the image processing pipeline, cell areas obtained using automated MATLAB-based segmentation were compared against manual measurements. For both lateral and radial–uniaxial stretching conditions, the percentage error between automated and manual measurements was calculated for representative cells before and after deformation (Table~\ref{table:S1} and Table~\ref{table:S2}). Across all cases, segmentation errors remained relatively low, supporting the use of the automated workflow for reliable quantification of cell geometry and deformation.

\begin{table}[h]
\centering
\caption{Verification of automated segmentation for lateral stretch (areas in pixels).}
\label{table:S1}
\begin{tabular}{|c|c|c|c|c|c|}
\hline
\multicolumn{3}{|c|}{\textbf{Before}} & \multicolumn{3}{c|}{\textbf{After}} \\
\hline
\textbf{MATLAB} & \textbf{Manual} & \textbf{\% Error} & \textbf{MATLAB} & \textbf{Manual} & \textbf{\% Error} \\
\hline
25436 & 26823 & 5 & 17450 & 17599 & 1 \\
19126 & 18593 & 3 & 16835 & 15422 & 9 \\
22700 & 23441 & 3 & 16575 & 17220 & 4 \\
17152 & 18592 & 8 & 10919 & 11499 & 5 \\
--    & --    & -- & 10354 & 10769 & 4 \\
--    & --    & -- & 3673  & 3754  & 2 \\
\hline
\end{tabular}
\end{table}

\begin{table}[h]
    \centering
\caption{Verification of automated segmentation for radial and uniaxial stretch (areas in pixels).}
\label{table:S2}
\begin{tabular}{|c|c|c|c|c|c|c|c|c|}
\hline
          \multicolumn{3}{|c|}{\textbf{Before}}& 
      \multicolumn{3}{|c|}{\textbf{After radial}} & \multicolumn{3}{|c|}{\textbf{After uniaxial}}\\ \hline
 MATLAB & Manual &  \%  $\Delta$  & MATLAB & Manual & \%  $\Delta$   & MATLAB & Manual & \%  $\Delta$  \\\hline
 31527& 31229& 1& 24625& 26395&7& 42282& 43226&2
\\
 25872& 26917& 4& 62747& 69687&10& 17780& 18900&6
\\
 25211& 26892& 6& 38060& 45936&17& 51824& 53915&4
\\
 5608& 6908& 19& 43870& 45391&3& 22859& 23126&1
\\
 6028& 6367& 5& & && & &
\\
 6651& 7702& 14& & && & &\\\hline\end{tabular}
\end{table}

\nolinenumbers

\end{document}